%% file: ms.tex
\documentclass[12pt,preprint]{aastex}

\newcommand{\oiv}{[\ion{O}{4}]}
\newcommand{\cii}{[\ion{C}{2}]}

\begin{document}

\title{Sub-Arcsecond Mid-Infrared Observations of NGC~6240: \\
Limitations of AGN-Starburst Power Diagnostics\altaffilmark{1} }

\author{E.\ Egami\altaffilmark{2,3}, G.\ Neugebauer\altaffilmark{2,3}, 
B.\ T.\ Soifer\altaffilmark{2,4}, K.\ Matthews\altaffilmark{2}, \\
E.\ E.\ Becklin\altaffilmark{5}, and M.~E.~Ressler\altaffilmark{6}}

\altaffiltext{1}{The data presented herein were obtained at
the W.\ M.\ Keck Observatory, which is operated as a scientific
partnership among the California Institute of Technology, the
University of California, and the National Aeronautics and Space
Administration.  The Observatory was made possible by the generous
financial support of the W.\ M.\ Keck Foundation.}

\altaffiltext{2}{Caltech Optical Observatories, California Institute
of Technology, 105-24, Pasadena, CA~91125}

\altaffiltext{3}{Present address: Steward Observatory, University of
Arizona, 933 N.\ Cherry Ave., Tucson, AZ 85721}

\altaffiltext{4}{Spitzer Science Center, California Institute of
Technology, 314-6, Pasadena, CA~91125}

\altaffiltext{5}{Department of Physics and Astronomy, University of
California, Los Angeles, CA90095-1562}

\altaffiltext{6}{Jet Propulsion Laboratory, California Institute of
Technology, 4800 Oak Grove Drive, Pasadena, CA~91109}

\begin{abstract}

In order to examine the relative importance of powerful starbursts and
Compton-thick AGNs in NGC~6240, we have obtained mid-infrared images
and low-resolution spectra of the galaxy with sub-arcsecond spatial
resolution using the Keck Telescopes.  Despite the high spatial
resolution ($\sim 200$ pc) of our data, no signature of the hidden
AGNs has been detected in the mid-infrared. The southern nucleus,
which we show provides 80--90\% of the total 8--25~$\mu$m luminosity
of the system, has a mid-infrared spectrum and a mid-/far-infrared
spectral energy distribution consistent with starbursts. At the same
time, however, it is also possible to attribute up to 60\% of the
bolometric luminosity to an AGN, consistent with X-ray observations,
if the AGN is heavily obscured and emits mostly in the far-infrared.
This ambiguity arises because the intrinsic variation of properties
among a given galaxy population (e.g., starbursts) introduces at least
a factor of a few uncertainty even into the most robust AGN-starburst
diagnostics.  We conclude that with present observations it is not
possible to determine the dominant power source in galaxies when AGN
and starburst luminosities are within a factor of a few of each other.

\end{abstract}

\section{INTRODUCTION}

NGC~6240 is a nearby (D $\sim$ 100 Mpc)\footnote{We adopt a distance
of 103.86 Mpc derived by \citet{Sanders03} based on a redshift of
0.0243 \citep{Solomon97} and the cosmic attractor model of
\citet{Mould00}.  The assumed cosmological parameters are
$\Omega_{M}=0.3$, $\Omega_{\Lambda}=0.7$, and $H_{0}=75$ km s$^{-1}$
Mpc$^{-1}$, respectively.  At this distance and redshift, 1\arcsec\
subtends 490 pc.}  system of interacting galaxies which epitomizes the
AGN-starburst controversy surrounding ultra-luminous infrared galaxies
(ULIRGs)\footnote{With an infrared luminosity $L_{IR}$ ($=
L$(8-1000~$\mu$m) as defined by \citet{Sanders96}) of
$7.1\times10^{11} L_{\sun}$ \citep{Sanders03}, NGC~6240 is a luminous
infrared galaxy (LIRG; $L_{IR} > 10^{11} L_{\sun}$) rather than an
ultra-luminous infrared galaxy (ULIRG; $L_{IR} > 10^{12} L_{\sun}$) in
the strict sense.  However, we do not make this distinction here
because NGC~6240 shares many of the ULIRG properties.}.  Together with
Arp~220 \citep{Soifer84}, NGC~6240 was one of the first few galaxies
discovered by IRAS to have an extremely large ($> 10^{11}$--$10^{12}
L_{\sun}$) and dominant ($> 0.9~L_{bol}$) infrared luminosity
\citep{Wright84}.  Its highly disturbed morphology \citep{Fosbury79}
and two nuclei separated by less than 2\arcsec\ ($< 1$ kpc)
\citep{Fried83} indicate that this is a system of two merging
galaxies.  Based on the shape of the infrared spectral energy
distribution (SED) and the large infrared size ($>$ 3 kpc) inferred
initially, it was argued that powerful starbursts induced by galaxy
interaction generate the large infrared luminosity by heating
interstellar dust with the UV radiation from young massive stars
\citep{Wright84,Joseph85}.

Although the size argument later turned out to be incorrect when the
mid-infrared size was measured to be $<$ 500 pc by \citet{Wynn93},
this starburst hypothesis for NGC~6240 has subsequently gained wide
support from a series of follow-up observations.  The most crucial
pieces of evidence are, (1) the deep CO absorption bandheads, (2) the
strong PAH features, and (3) the starburst-like radio-infrared
luminosity ratio.  The deep CO absorption bandheads detected in the
near-infrared clearly indicate that the near-infrared continuum of
NGC~6240 is dominated by starlight
\citep{Rieke85,Lester88,Ridgway94,Shier96,Sugai97,Tecza00}.  The most
recent study by \citet{Tecza00} shows that these CO absorption
bandheads are likely to be produced by late K or early M supergiants,
and that the starburst population associated with these supergiants,
estimated to be triggered $\sim2\times10^{7}$ years ago and lasting
$\sim5\times10^{6}$ years, are powerful enough to provide 30--100\% of
the bolometric luminosity.  The strong PAH features seen in the
mid-infrared spectra of NGC~6240
\citep{Smith89,Genzel98,Dudley99,Rigopoulou99,Imanishi00,Lutz03} also suggest
starbursts as the main luminosity source since they are much weaker in
Active Galactic Nuclei (AGN) galaxies.  Finally, the radio-infrared
luminosity ratio of NGC~6240, averaged over the whole galaxy, is also
starburst-like \citep{Colbert94,Tecza00} although the existence of
compact radio cores ($< 26$ pc) in the two nuclei may suggest some AGN
radio emission \citep{Colbert94,Beswick01,Gallimore04}.\ Together with
the starburst-driven superwind seen both in the visual
\citep{Heckman87,Armus90,Heckman90} and in the soft X-ray
\citep{Schulz98,Komossa98,Iwasawa98,Lira02}, the evidence for the
powerful starbursts is indeed strong.

This starburst picture was seriously challenged by a series of the
strong hard X-ray detections, which suggest the existence of a
QSO-like AGN in NGC~6240.  An initial hint came from the ASCA hard
X-ray spectrum, which showed an extremely flat continuum above 3~keV
with a strong iron K line complex
\citep{Mitsuda95,Kii97,Iwasawa98,Nakagawa99}.  This suggests that what
we are seeing is reflected X-ray emission from an AGN obscured by a
Compton-thick ($N_{H}>~2\times10^{24}$cm$^{-2}$) material, and that
the intrinsic luminosity of this AGN could be as high as
$2.6\times10^{11}~L_{\sun}$($10^{45}$~erg~s$^{-1}$) \citep{Iwasawa98}.
This conclusion was later confirmed by the Beppo-SAX observations,
which showed that the X-ray spectrum of NGC~6240 clearly exhibits the
emergence of a powerful hard X-ray continuum above 10~keV from the
obscured AGN \citep{Vignati99}.  If we take the estimated intrinsic
2--10~keV nuclear luminosity of $>~10^{44}$ erg~s$^{-1}$
(absorption-corrected based on the Beppo-SAX spectrum extending up to
100 keV) and assume that it is less than 10\% of the AGN bolometric
luminosity ($L_{2-10 keV}/L_{bol} \sim 0.03-0.1$ for Seyferts and QSOs
according to \citet{Iwasawa01}), this sets a lower limit of
$10^{45}$~erg~s$^{-1}$ on the AGN luminosity in NGC~6240, which is
consistent with the ASCA result. Given various uncertainties, the AGN
can also produce 50--100\% of the bolometric luminosity of NGC~6240.
Recently, Chandra hard X-ray images of NGC~6240 have shown that this
hard X-ray emission originates from two nuclei \citep{Komossa03}.  A
detailed X-ray spectroscopic study has also been performed by
\citet{Boller03} and \citet{Netzer05} based on the XMM-Newton X-ray data.

In the last few years, NGC~6240 has been studied with increasingly
high spatial resolution.  \cite{Gerssen04} resolved both nuclei into
separate components using HST observations at visual wavelengths, but
they find no clear sign of the two AGNs.  They emphasize, however, the
importance of X-ray data for identifying AGNs in highly
dust-enshrouded environments. Very Long Baseline Array (VLBA)
observations at radio frequencies reveal three compact sources in
NGC~6240 \citep{Gallimore04}, two of which are associated with the
X-ray nuclei.

The two conflicting pictures put NGC~6240 at the heart of the
AGN-starburst controversy within ULIRGs: that is, do AGNs or
starbursts dominate their luminosity output?  In recent years, the
prevailing view has been that it is starbursts that dominate.  The
majority of ULIRGs shows no obvious sign of a hidden AGN, and the
observed starburst components seem powerful enough to provide the
bolometric luminosity of these galaxies when corrected for extinction
\citep{Genzel98}.  However, NGC~6240 is a notable anomaly in this
picture with strong evidence for a dust-obscured powerful AGN.  If
this AGN indeed provides a substantial fraction of the infrared
luminosity, the validity of various AGN-starbursts power diagnostics
must be re-examined since many of these diagnostics indicate NGC~6240
to be a typical starburst galaxy.  Most recently, the existence of a
Compton-thick AGN was also hinted for Arp 220, another ULIRG with no
sign of AGN signature in the infrared, through the detection of the Fe
K emission in the hard X-ray \citep{Iwasawa05}.  The SED modeling of
Arp~220 by \citet{Spoon04} suggests that such a heavily obscured AGN
might contribute significantly to the bolometric luminosity.

Here we present mid-infrared images and spectra of NGC~6240 obtained
with the Keck Telescopes with sub-arcsecond spatial resolutions.  The
Keck Telescopes routinely deliver a diffraction-limited spatial
resolution of 0\farcs3--0\farcs5 in the mid-infrared, which
corresponds to 150--250~pc at the distance of NGC~6240.  Such a high
spatial resolution is essential to probe a complicated merging system
like NGC~6240, in which the two nuclei are separated by only
1\farcs4--1\farcs5 (1\farcs5 in VLBA --- \citep{Gallimore04}; 1\farcs4
in X-rays --- \citep{Komossa03}).

\section{OBSERVATIONS AND DATA REDUCTION}

\subsection{Keck~II/MIRLIN Mid-Infrared Images}

The mid-infrared images of NGC~6240 were taken on UT 1998 March 19
with the MIRLIN camera \citep{Ressler94} on the Keck~II 10m Telescope
on Mauna Kea in Hawaii.  The camera uses a 128$\times$128 Si:As BIB
array, and was attached to the f/40 bent Cassegrain visitor port,
producing a pixel scale of 0\farcs138 pixel$^{-1}$ with a field of
view 18\arcsec\ on a side.  Secondary-mirror chopping and telescope
nodding were employed for effective subtraction of the sky background
and instrumental noise.  The observational methods and characteristics
as well as the photometric calibration and data reduction are the same
as those used by \citet{Soifer99,Soifer00} and are described therein.

\begin{figure}
  \vspace*{-3cm}

  \plotone{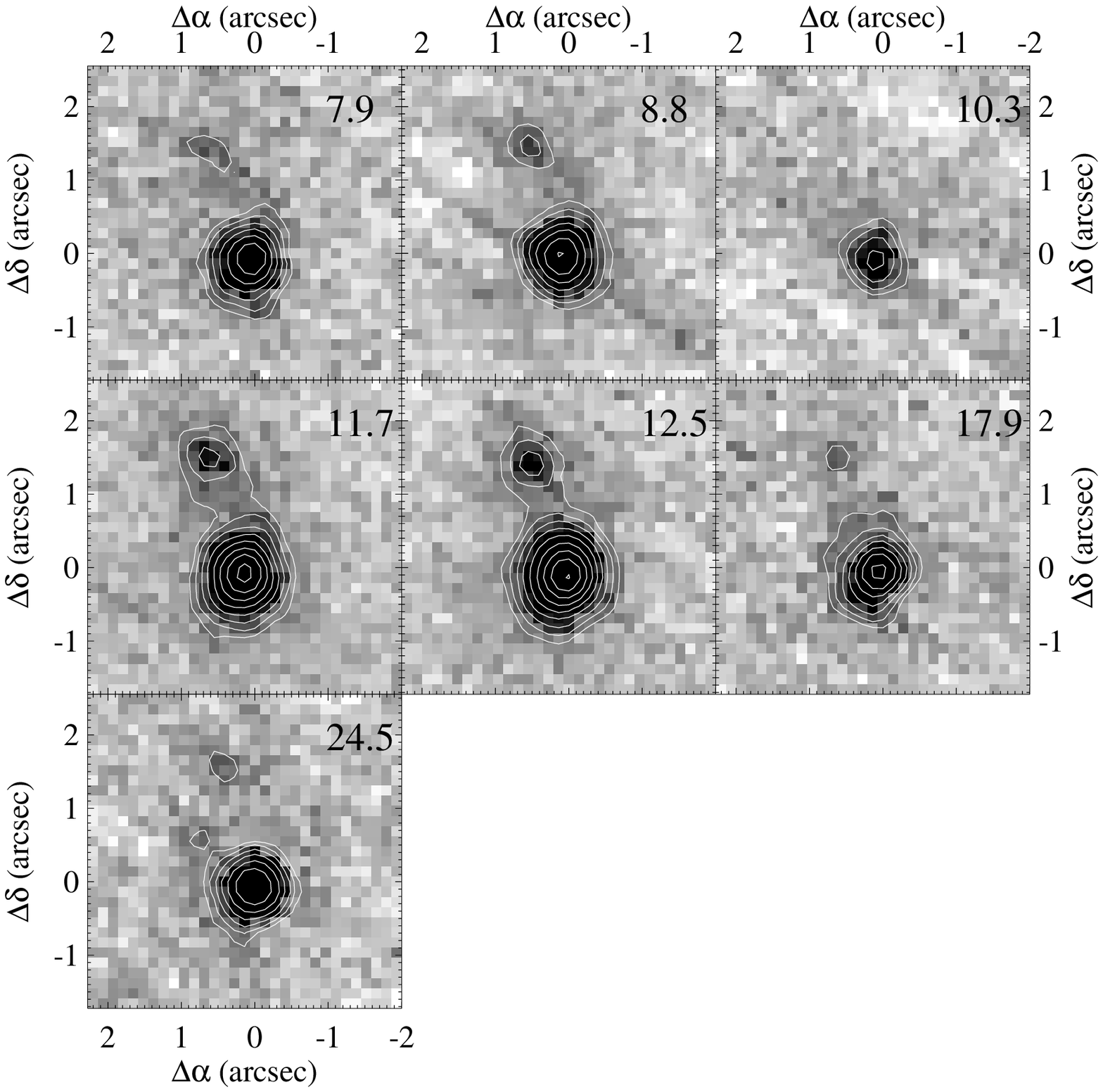}

  \vspace*{-2cm}

  \caption[f1.eps]{Mid-infrared images of NGC~6240. North is up
  and east is left.  The observed wavelength (in~$\mu$m) is denoted in
  each panel.  The contour maps were produced from Gaussian-smoothed
  images (Gaussian FWHM $=$ 2 pixels).  The lowest contour corresponds
  to 2$\sigma$ above the sky level, which was calculated in the
  non-smoothed image, and the successive contours correspond to a
  factor of 1.4 increases in surface brightness.  The diagonal lines
  from upper left to lower right seen in some images (e.g., at 8.8
  ~$\mu$m) are due to the noise pattern of the array.  \label{mirlin}}

\end{figure}

Observations of NGC~6240 were made in the seven mid-infrared filters
listed in \citet{Soifer99}, where the central wavelengths and filter
full widths are given.  The central wavelengths are also listed in
Table~\ref{flux} together with the on-source integration times.  The
mean airmass was 1.07.\ The MIRLIN observations emphasized photometry
rather than spatial structure, and as a result no strict size limit
can be placed on the observations of the nucleus of NGC~6240.\ In
particular, since no star was observed close in time, we do not have
accurate estimates of the point spread function (PSF) and seeing.
However, the bright compact southern nucleus of NGC~6240
(Figure~\ref{mirlin}) has FWHMs of 0\farcs5--0\farcs7 at
7.9--24.5~$\mu$m, indicating that the intrinsic seeing was at least
this good.

\subsection{Keck~I/LWS Mid-Infrared Spectra}

The low-resolution ($\Delta \lambda / \lambda \sim 50$) mid-infrared
spectra of NGC~6240 were obtained with the Long Wavelength
Spectrometer (LWS; \citet{Jones93}) on the Keck~I 10m Telescope.  LWS
uses a 128$\times$128 BIB array, and is attached to the f/25 forward
Cassegrain focus of the telescope, producing a pixel scale of
0\farcs08 pixel$^{-1}$ with a field of view 10\arcsec\ on a side.  The
slit width was 0\farcs48 (6 pixels).  The observing methods,
instrumental characteristics, and data reduction used to obtain the
mid-infrared spectra are the same as were used by \citet{Soifer02},
and are described therein.

The spectra of NGC~6240 covering 8.8$~\mu$m--13.1$~\mu$m were taken on
UT 2000 May 21 while those covering 7.5~$\mu$m--12.1$~\mu$m were taken
on UT 2000 May 22.  The total integration times are 147 minutes and 59
minutes, respectively.  The slit position angle was 16\degr\ (east of
north), which is along the direction connecting the two nuclei. The
NGC~6240 spectra were divided by the spectra of HR~5340 and HR~6406,
and multiplied by a blackbody spectrum at the star's effective
temperature.  For wavelength calibration, we adopted the spectral
dispersion of 0.0375~$\mu$m~pixel$^{-1}$ derived from our other LWS
data, and adjusted the offset by using the PAH 11.3~$\mu$m feature
seen redshifted to 11.6~$\mu$m in the NGC~6240 spectra.

\section{RESULTS}

\subsection{Mid-Infrared Morphology and Photometry}

Figure~\ref{mirlin} shows the mid-infrared images.  The images clearly
show two compact nuclei separated by $\sim$ 1\farcs6~$\pm~$0\farcs1
along the position angle of 16\degr ~(east of north).  The southern
nucleus dominates the mid-infrared luminosity output.  It appears
round and symmetric although some north-south elongation is seen in
the 11.7 and 12.5~$\mu$m images, which are of the highest
signal-to-noise ratio. The sub-components of the two main sources seen
by \citet{Max05} in the near infrared and by \citet{Gallimore04} with
the VLBA cannot be accurately resolved with the present mid-infrared
resolution, but the elongations in the mid-infrared images hint at
this structure.

Ignoring the elongation, we have estimated the source size of the
southern nucleus by fitting a circularly symmetric Gaussian to the
observations.\ Since we do not have the data to determine the seeing
at the time of the NGC~6240 observations, we simply subtracted, in
quadrature, the diffraction-limited beam sizes from the measured
Gaussian FWHMs and derived the most indicative upper limits on the
intrinsic size.\ There was, however, no evidence that its size at any
of the mid-infrared wavelengths exceeded 200~pc (0\farcs4) after
diffraction broadening of the images is accounted for.

Table~\ref{flux} lists the continuum flux densities of the two nuclei
measured in a circular 1\arcsec\ diameter beam as well as the total
flux densities for the whole system measured in a 4\arcsec\ diameter
circular beam.  The total flux densities at 12.5 and 24.5~$\mu$m
measured by MIRLIN are 0.5 and 3.3 Jy, respectively, which are close
to the IRAS flux densities of 0.6 and 3.4 Jy at 12 and 25~$\mu$m.
This indicates that these two compact nuclei provide almost all the
mid-infrared luminosity of NGC~6240.  The southern nucleus is
especially luminous, providing $\sim$ 80--90\% of the total flux at
these wavelengths.  

Table~\ref{fratio} compares the flux ratio of the two nuclei at a
number of wavelengths from X-ray to radio.  The flux ratio increases
by a factor of two from near-infrared to mid-infrared, indicating that
the southern nucleus is slightly redderr than the northern one in this
wavelength range.  However, the ratios are quite similar in the X-ray
and radio ($\sim$2--3).  

\begin{figure}
  \vspace*{-1cm}

  \epsscale{0.7}

  \plotone{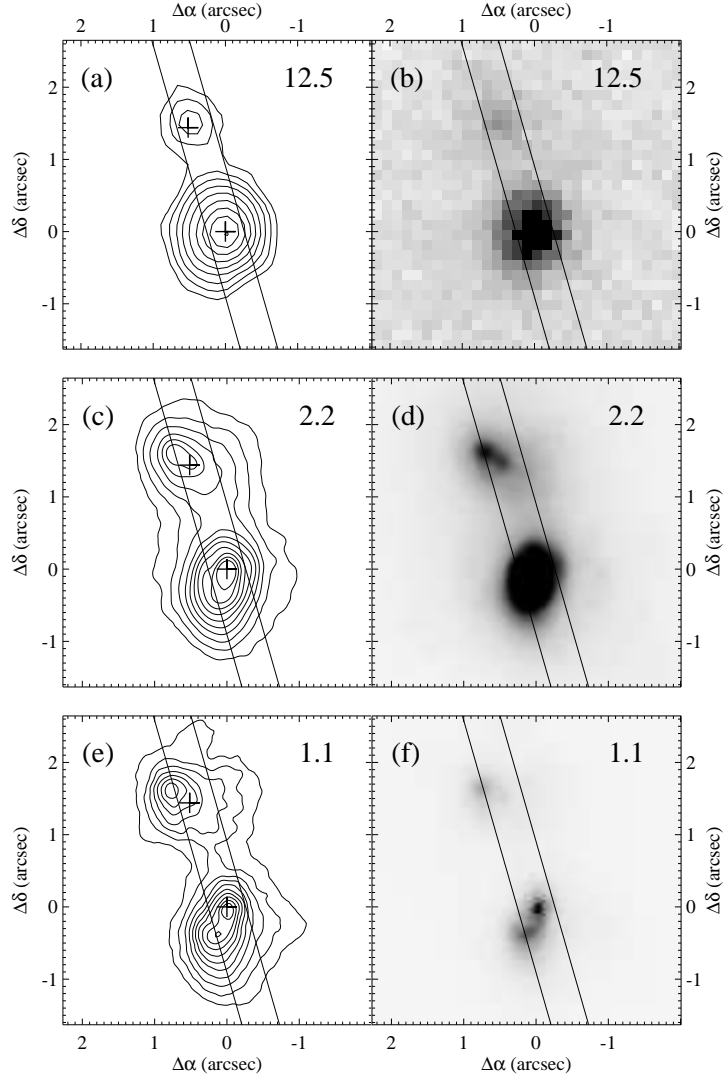}

  \vspace*{0cm}

  \caption[f2.eps]{The 12.5~$\mu$m image (top row) compared with
 the HST/NICMOS images at 2.2~$\mu$m (middle row) and 1.1~$\mu$m
 (bottom row) from \citet{Scoville00}.  North is up and east is left.
 In each row, the left panel shows the contour map while the right
 panels show the image itself with different gray scales to emphasize
 the two sources in each nucleus.  The contour maps were produced from
 Gaussian-smoothed images (Gaussian FWHM $=$ 2 pixels).  The lowest
 contour corresponds to 2$\sigma$ above the sky level, which was
 calculated in the non-smoothed image, and the successive contours
 correspond to a factor of 1.4 increases in surface brightness.  The
 two crosses overlaid on the contour plots indicate the positions of
 the two radio peaks \citep{Gallimore04} with the assumption that the
 southern nucleus is spatially coincident at radio and mid-infrared
 wavelengths. The slit position of the LWS spectra, which has a
 position angle of 16\degr\ (east of north), is also overlaid.
 \label{nicmos}}

\end{figure}

Figure~\ref{nicmos} compares the MIRLIN 12.5~$\mu$m image, which has
the highest signal-to-noise ratio, with the HST/NICMOS 1.1 and
2.2$~\mu$m images \citep{Scoville00}.  The positions of the two radio
nuclei \citep{Gallimore04} are also overlaid (crosses) with the
assumption that the positions of the southern nucleus are spatially
coincident. Note that it is not possible to perform independent
astrometric calibration with the mid-infrared images because NGC~6240
is the only detected source in the small field of view of the MIRLIN
images.  As the figure shows, the separation and position angle of the
two nuclei at 12.5~$\mu$m are almost exactly the same as those at
radio, and hence X-ray, wavelengths.  This suggests that we are
looking at the same components in the mid-infrared as we see at the
other wavelengths.

The HST/NICMOS images, on the other hand, show a slightly larger
separation between the nuclei, and the northern nucleus is
significantly displaced from the radio position.\ In fact,
\citet{Gerssen04} have identified a secondary peak in the HST/NICMOS
images of the northern nucleus of NGC~6240 close to the position of
the radio source.\ Its faintness in the 1.1~$\mu$m image
(Figure~\ref{nicmos}f) indicates that this source is likely to be
heavily extincted by dust absorption in the galaxy.\ The positional
coincidence argues that this fainter secondary component is
responsible for the mid-infrared and radio luminosity of the northern
nucleus.\ The southern nucleus also displays a substantial north-south
elongation in the HST/NICMOS images due to another secondary component
seen conspicuously in the 1.1~$\mu$m image (Figure~\ref{nicmos}f).

\subsection{Mid-Infrared Spectra}

\begin{figure}

  \vspace*{-2.5cm}

  \epsscale{1}

  \plotone{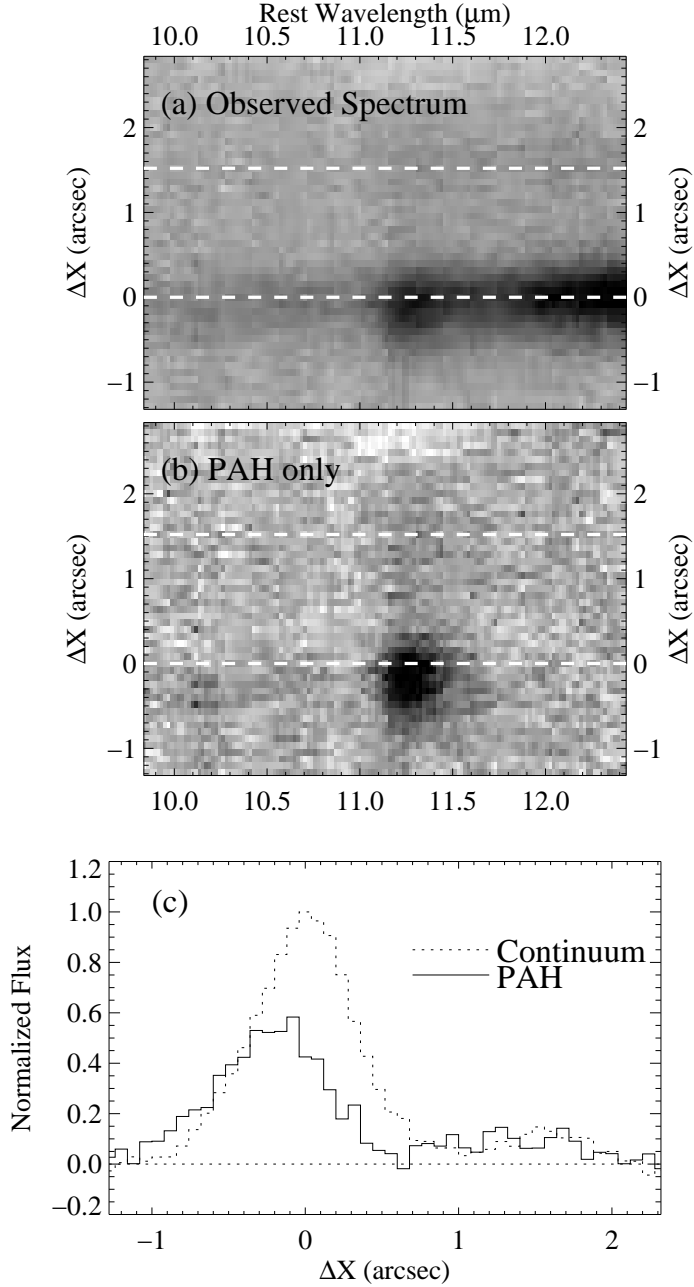}

  \vspace*{-2cm}

  \caption[f3.eps]{The two-dimensional LWS spectrum of NGC~6240
  covering 9.9--12.4~$\mu$m in the restframe: (a) the observed
  spectrum; (b) the continuum-subtracted spectrum, showing a strong
  11.3~$\mu$m PAH feature in the southern nucleus; (c) the spatial
  profiles of the continuum (dotted line) and 11.3~$\mu$m PAH feature
  (solid line) produced by summing the 11.0--11.7~$\mu$m
  region. Normalization of both components was achieved by making the
  integration of the continuum peak of the southern nucleus equal to
  unity.\ The two horizontal white dashed lines in (a) and (b)
  indicate the positions of the northern (upper) and southern (lower)
  nucleus, respectively although the northern nucleus is too faint to
  see.  \label{lws_long}}

\end{figure}

Figure~\ref{lws_long}a shows the LWS two-dimensional spectrum
(wavelength vs. location along the slit) of NGC~6240 covering the
wavelength range from 9.9~$\mu$m to 12.4~$\mu$m in the restframe.  The
two spectra have been combined to form this two-dimensional spectrum.
The actual slit position is shown in Figure~\ref{nicmos}.  The
southern nucleus is clearly visible, but the northern nucleus at the
position of the upper white dashed line is too faint to see.  We
therefore exclude the northern nucleus from the following discussion.
The southern nucleus has a detectable but faint continuum at $\lesssim
11~\mu$m, which rises sharply toward longer wavelengths.  The
prominent bright spot between 11 and 11.5~$\mu$m is due to the
11.3~$\mu$m PAH feature.

Figure~\ref{lws_long}b shows the same spectral image with the
continuum subtracted from the southern nucleus.  The continuum
subtraction was done in the manner described in \citet{Soifer02}.  The
11.3~$\mu$m PAH feature is clearly seen.  However, the line emitting
region seems to be displaced from the continuum by 0\farcs2 ($\sim$90
pc) downward, which corresponds to the south-west direction.  This can
be clearly seen when the spatial profiles of the continuum and PAH
feature are compared (Figure~\ref{lws_long}c).\ Incidentally, this
separation is similar to that of the two near-infrared peaks in the
southern nucleus seen in the {\em HST}/NICMOS 1.1 $\mu$m image
(Figure~\ref{nicmos}e and f), suggesting a possible connection between
the PAH peak and the secondary near-infrared peak.  We also note that
there is no obvious detection of the PAH emission from the molecular
gas concentration between the two nuclei detected with the
near-infrared H$_{2}$ emission
\citep{Herbst90,Werf93,Sugai97,Tecza00,Bogdanovic03} and with the
millimeter CO emission \citep{Tacconi99}.

\begin{figure}
  \epsscale{0.8}

  \plotone{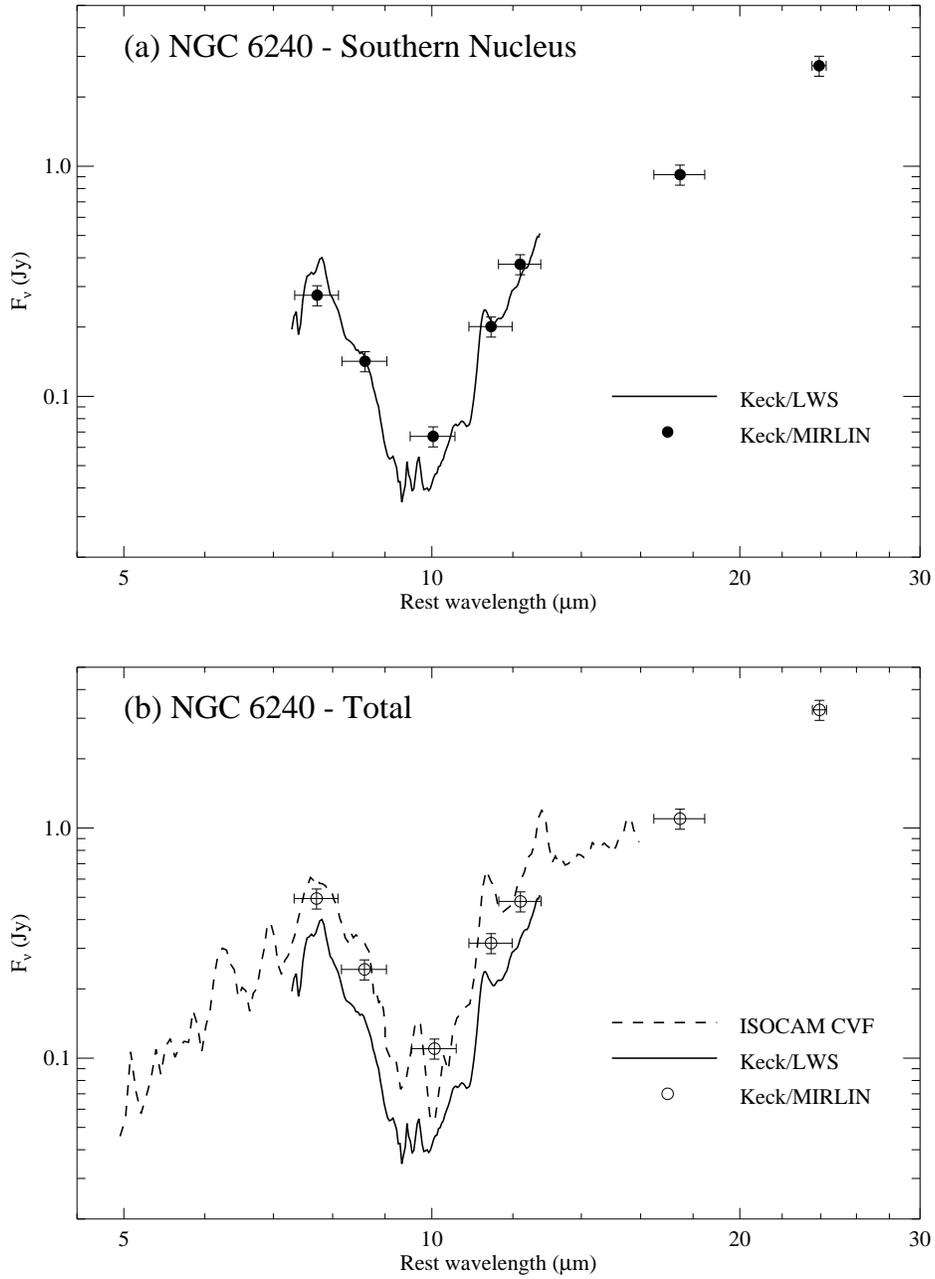}

  \caption[f4.eps]{(a) The LWS spectrum (solid line) and MIRLIN
  photometric measurements (solid circles) of the southern nucleus in
  NGC~6240; (b) The ISOCAM CVF spectrum by \citet{Lutz03} (dashed
  line) and MIRLIN photometric measurements with a 4\arcsec\ diameter
  beam (open circles).  For comparison, the LWS spectrum of the
  southern nucleus shown in (a) is over-plotted (solid line).  The
  horizontal bars with the MIRLIN points indicate the photometric
  passbands while the vertical bars show a conservative photometric
  uncertainty of $\pm$10\%.
  \label{lws}}
\end{figure}

Figure~\ref{lws}a shows the LWS spectrum of the southern nucleus over
the full wavelength range from 7.3 to 12.8~$\mu$m in the rest frame,
together with the photometric measurements with MIRLIN. To produce the
one dimensional spectrum, the signals within 7 pixels (0\farcs56) from
the continuum center were coadded.  This extraction aperture is large
enough to contain most of the continuum and PAH emission, and
therefore the 0\farcs2 offset of the PAH emitting region mentioned
above does not affect the resultant spectrum significantly (the effect
on the PAH flux is a $\sim$ 5\% reduction at most).  The flux level of
the spectrum was adjusted such that it makes a reasonable match to the
broad band MIRLIN photometric points.  A strong PAH 11.3~$\mu$m
feature is seen, and its line flux is measured to be
$8.6\times10^{-16}$ W m$^{-2}$. Also, the sharp drop of flux at the
short-wavelength end of the spectrum suggests that we are seeing the
peak of the 7.7~$\mu$m PAH feature while the shoulder to the right
indicates the existence of the 8.6~$\mu$m PAH feature.

For comparison, the ISOCAM CVF spectrum presented by \citet{Lutz03} is
shown in Figure~\ref{lws}b. (\citet{Lutz03} also give ISOSWS data, but
these are high resolution spectra targeted on individual lines).
Since the ISOCAM CVF spectrum contains the fluxes from both nuclei, it
is compared with the total flux measurements by MIRLIN with a
4\arcsec~diameter circular beam.  The two data sets show good
agreement, confirming the accuracy of the photometric calibration. The
LWS spectrum of the southern nucleus is over-plotted, showing that the
overall continuum shape (e.g., the depth of the 10 $\mu$m trough) is
quite similar to that of the ISOCAM spectrum although the emission
features (e.g., 11.3 $\mu$m PAH) are much weaker.

\section{DISCUSSION}
\subsection{The Southern Nucleus}

As is the case with other ULIRGs \citep{Soifer00}, the southern
nucleus of NGC~6240, which dominates the luminosity output, is
characterized by an extremely high surface brightness.  If we take the
intrinsic source area to be a circular disk with a diameter of 200 pc
on the sky, the bolometric luminosity of $7\times10^{11}~L_{\sun}$
translates into a surface brightness of
$2\times10^{13}~L_{\sun}$~kpc$^{-2}$.  This is at the high end of
surface brightnesses of typical starburst galaxies
\citep{Soifer00,Evans03}. The real surface brightness is probably even
higher since we have taken an upper limit on the size of the continuum
emitting region.

\begin{figure}

  \hspace*{0.5cm}\includegraphics[angle=90,scale=0.7]{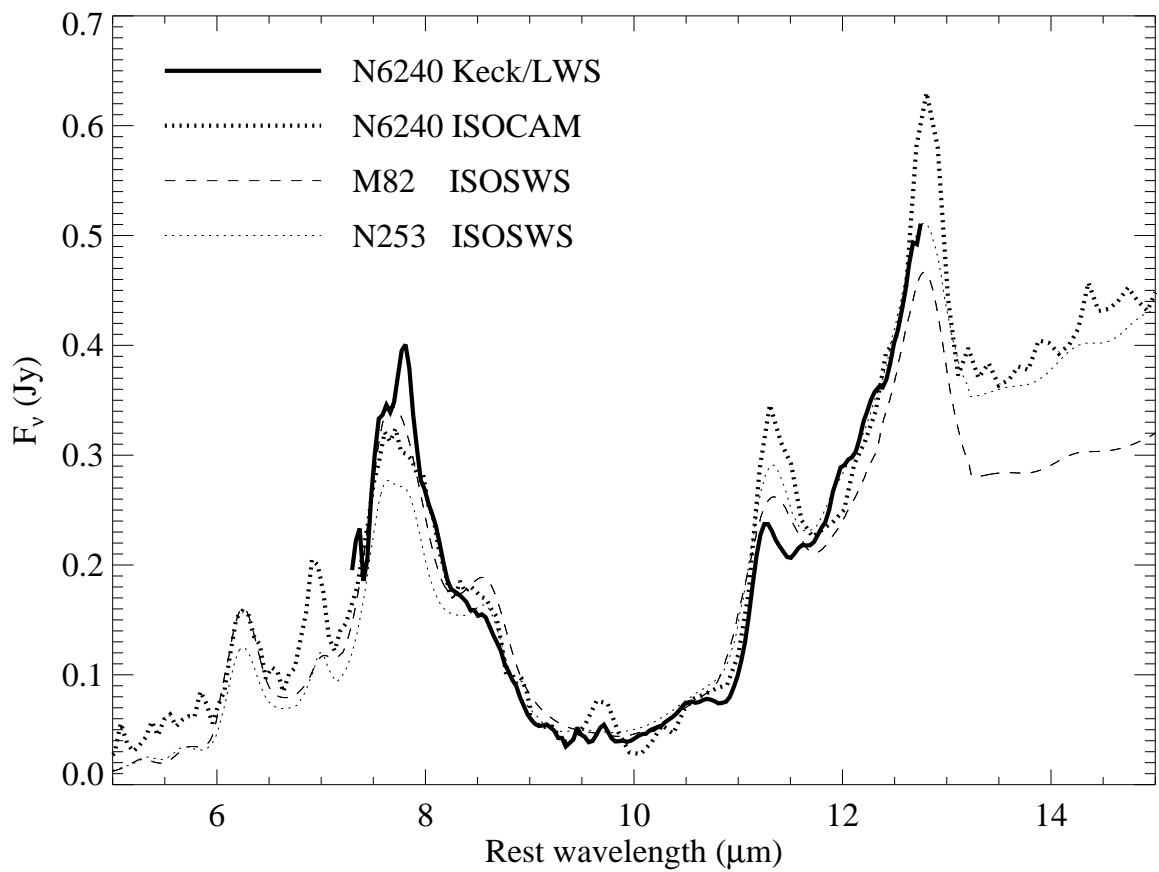}

  \caption[f5.eps]{The mid-infrared spectra of the southern
  nucleus of NGC~6240 are compared with those of the local starburst
  galaxies M~82 and NGC~253.  The thick solid line is the Keck/LWS
  mid-infrared spectrum of the southern nucleus while the thick dotted
  line is the ISOCAM CVF spectrum \citep{Lutz03} scaled to match the
  LWS spectrum. Note that the ISOCAM spectrum contains both nuclei and
  the surrounding area.  The thin dashed and dotted lines show the
  ISOSWS spectra of M~82 and NGC~253 \citep{Sturm00} scaled to match
  the LWS spectrum.
  \label{spmodela}}

\end{figure}

Despite this high surface brightness, the mid-infrared spectrum of the
southern nucleus is similar to those of the much less luminous dusty
starburst galaxies such as M~82 and NGC~253.  Figure~\ref{spmodela}
compares the LWS spectrum of the southern nucleus with the ISOSWS
spectra of M~82 and NGC~253 \citep{Sturm00}.  These spectra can be
compared directly since they both sample the light within an area of a
few hundred parsec in diameter around the nucleus.  The match is
reasonable in terms of the 10~$\mu$m trough depth and PAH feature
strengths.  On the basis of the similarity of the mid-infrared
spectrum to those of M~82 and NGC~253, we would argue that the
mid-infrared emission originates from starbursts.

The good agreement between the LWS and ISOCAM spectra seen in
Figure~\ref{spmodela} can be interpreted as a consequence of the
southern nucleus dominating the total luminosity of the system.  For
this reason, we will not distinguish between the LWS spectrum and the
ISOCAM spectrum in the following discussion once an appropriate
scaling is applied to the latter to match the former.

\begin{figure}

  \hspace*{0.5cm}\includegraphics[angle=90,scale=0.7]{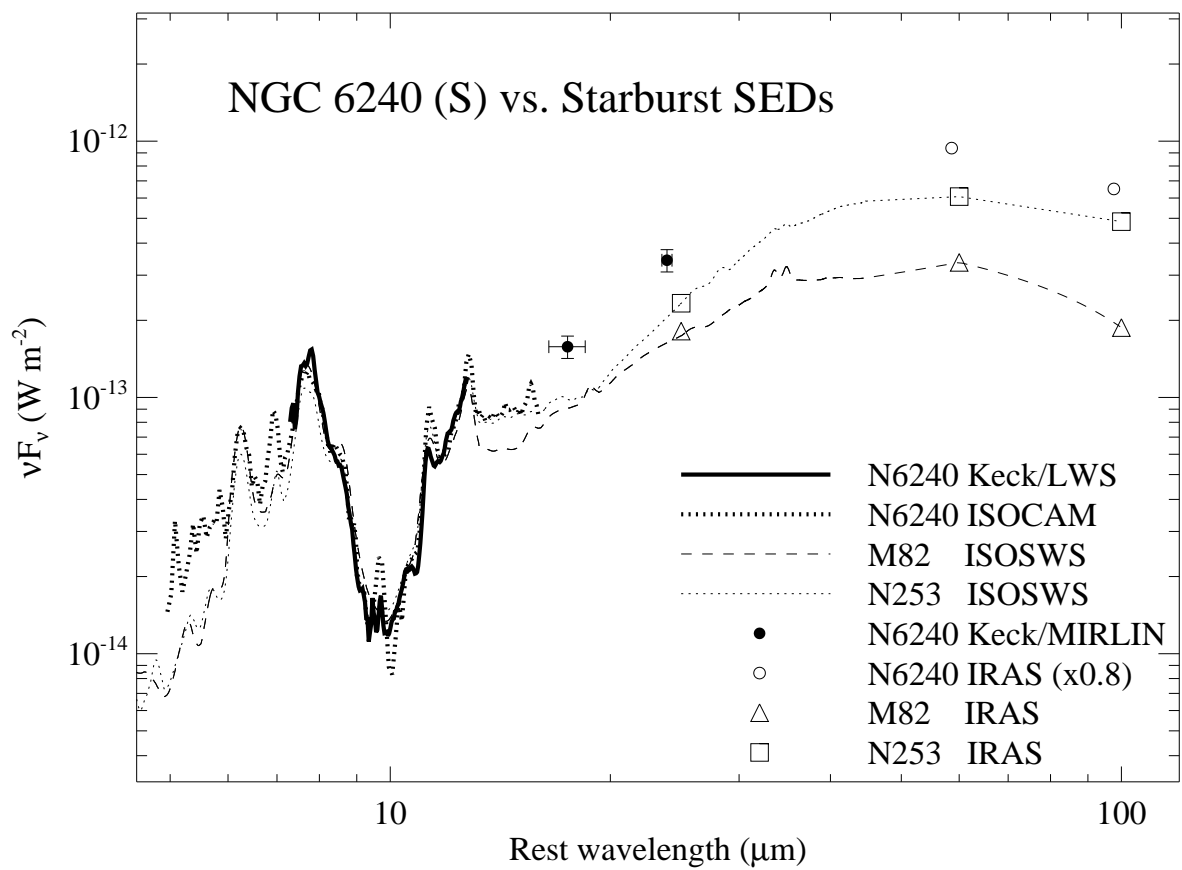}

  \caption[f6.eps]{The mid-/far-infrared SEDs of the southern
  nucleus of NGC~6240 are compared with those of the local starburst
  galaxies M~82 and NGC~253.\  The ISOSWS spectra were extended to
  100~$\mu$m using the IRAS 25, 60 and 100~$\mu$m measurements
  (triangles and squares) scaled such that the IRAS and ISOSWS agree
  at 25~$\mu$m.  The 17.9 and 24.5~$\mu$m photometric points of the
  NGC~6240 southern nucleus are from MIRLIN while those at 60 and
  100~$\mu$m are from IRAS with a scaling of $\times0.8$ to account
  for the southern nucleus contribution to the total flux.
  \label{spmodelb}}

\end{figure}

Although the 7--13~$\mu$m spectra match well, the spectrum/SED of the
NGC~6240 southern nucleus outside this range is different from those
of M~82 and NGC~253 in three respects (Figure~\ref{spmodelb}): (1) the
NGC~6240 SED rises more sharply in the mid-infrared; (2) the NGC~6240
spectrum drops more slowly at $<~7~\mu$m; and (3) the NGC~6240 SED has
more luminosity in the far-infrared.  The third point is less
certain because this depends on the far-infrared SED of the southern
nucleus, which we cannot measure directly.  In Figure~\ref{spmodelb},
we estimated the southern nucleus far-infrared SED of NGC~6240 from
the IRAS 60/100 $\mu$m measurements assuming a luminosity ratio of 4
between the two nuclei (i.e., the southern nucleus luminosity
contribution is 80\%).  If the true ratio is larger as suggested by
the mid-infrared measurements (up to $\sim$8), then the discrepancy
becomes larger; On the other hand, if the true ratio is as small as
the radio measurements indicate (down to $\sim$ 2), the disrepancy, at
least in comparison with NGC~253, would go away.

It is interesting that between the two SEDs of the starburst galaxies,
the SED of NGC~253 is significantly closer to that of NGC~6240.
NGC~253 is also known to contain an AGN, detected in the radio
\citep{Ulvestad97} and X-ray \citep{Weaver02}, and this might explain
the similarity.  In fact, as we shall show below, it is possible to
reproduce the SED of NGC~6240 from that of M~82 by adding spectral
components produced by a heavily dust-obscured AGN.  On the other
hand, the AGN in NGC~253 is thought to be much weaker than that in
NGC~6240, and therefore it may not have a significant effect on the
shape of the infrared SED.  In this case, the difference between the
SEDs of M~82 and NGC~253, though quite significant
(Figure~\ref{spmodelb}), should simply be understood as a variation of
properties among starburst galaxies.  Although the situation is not yet
clear, the simplest interpretation would be the latter: that is, there
is a significant variation of properties among starburst galaxies, and
therefore that everything we see in the infrared spectrum/SED of
NGC~6240 can still be explained by starbursts.

\subsection{The Hidden AGN}

Given the fact that the AGN in NGC~6240 is hidden behind a
Compton-thick wall of material \citep{Vignati99}, it is not surprising
that there is no clear AGN signature seen in the mid-infrared.  A
column of Compton-thick material ($N_{H} > 2\times10^{24}$~cm$^{-2}$)
could produce a visual extinction of $A_{V} > $~1000 mag, which is
simply too large to be penetrated in the mid-infrared.  In fact,
\citet{Krabbe01} have shown that compared with Seyfert galaxies,
NGC~6240 is an order of magnitude underluminous in the mid-infrared
with respect to its hard X-ray luminosity, which suggests that the AGN
light is highly absorbed even in the mid-infrared.

\begin{figure}

  \hspace*{0.5cm}\includegraphics[angle=90,scale=0.7]{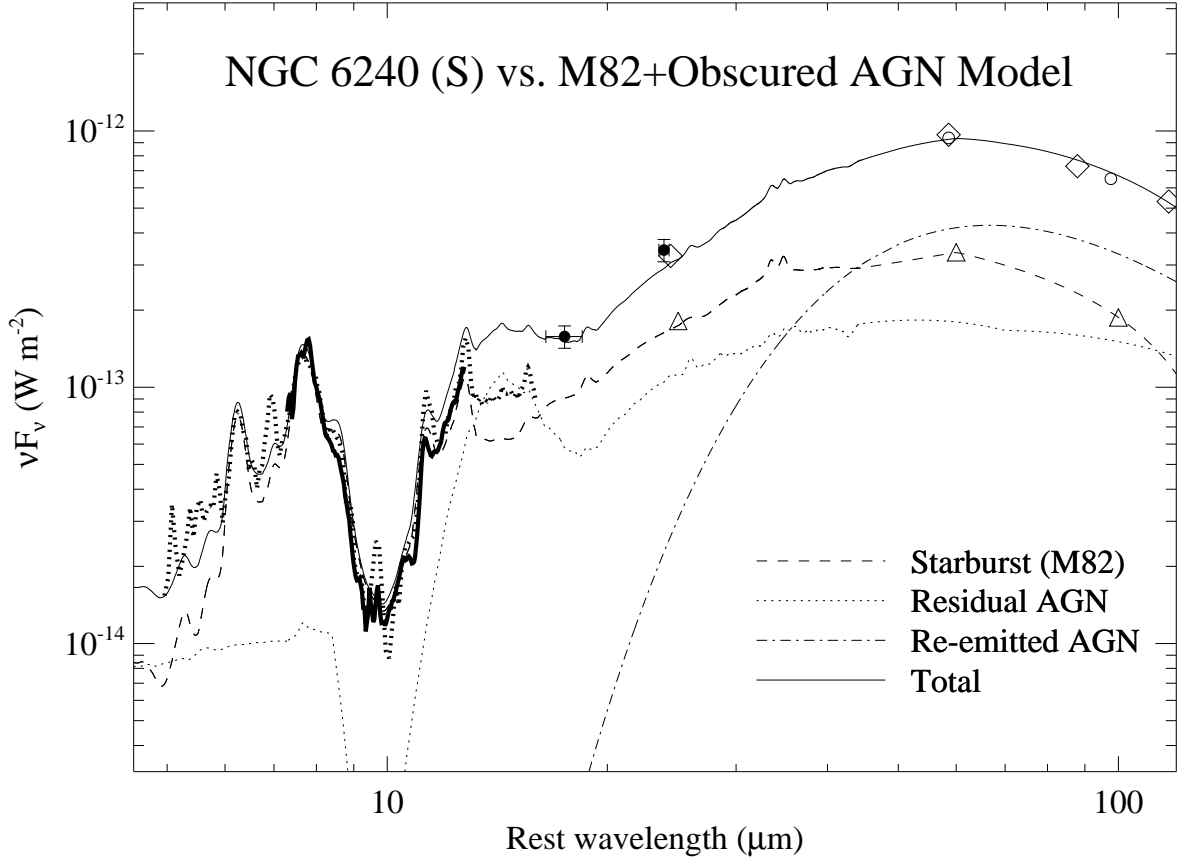}

  \caption[f7.eps]{SED model fit for the southern nucleus with
  both starburst and AGN components.\ The observed mid-infrared
  spectra and mid-/far-infrared SED shown in Figure~\ref{spmodelb} are
  reproduced by three components as discussed in the text: (1)
  starburst --- M~82 spectrum (dashed line; open triangles); (2) the
  residual AGN emission after absorption --- NGC~1068 spectrum
  reddened by $A_{V}=70$~mag (dotted line); (3) re-emitted AGN ---
  55~K black-body radiation (dash-dot line).\ The sum of these three
  components is shown by the thin solid line.\ The diamonds are the
  measurements by \citet{Klaas01}, and were scaled by
  $\times0.8$ to simulate the fluxes from the southern nucleus.
  The open and solid circles are the NGC~6240 photometric measurements
  shown in Figure~\ref{spmodelb}.
  \label{spmodel2}}
\end{figure}

We have estimated the maximum AGN contribution allowed by the data
using a simple model. The mid-infrared radiation of the AGN was
assumed to be heavily extincted so as to erase the AGN spectrum in the
mid-infrared, and the absorbed mid-infrared luminosity was assumed to
be re-emitted in the far-infrared as thermal radiation from dust. The
emission resulting from this model is shown in Figure~\ref{spmodel2}
and consists of the following three components: (1) starburst
emission, (2) residual AGN emission after absorption, and (3)
re-emitted AGN emission.  The M~82 SED shown in Figure~\ref{spmodelb}
was used to represent the starburst.\ A spectrum of NGC~1068
\citep{Sturm00} reddened by $A_{V}=70$~mag represented the residual
AGN emission.\ The NGC~1068 spectrum was extended to longer
wavelengths by extrapolating the power-law continuum at
18--45~$\mu$m. This extrapolation does not affect the modeling results
as long as it is much less luminous than the other components in the
far-infrared.\ The re-emitted AGN was represented by a black-body
emission with a temperature of 55 K, a value that produces a good fit
to the observed SED (a similar temperature of 57 K was also derived by
\citet{Klaas01}).  In this particular model, the luminosity of the
starburst in the restricted 4.5--120~$\mu$m range is
$1.8\times10^{11}~L_{\sun}$, that emitted by the AGN after extinction
is $1.1\times10^{11}~L_{\sun}$ and that of the re-emitted AGN
$1.5\times10^{11}~L_{\sun}$, which makes the total emitted luminosity
ascribable to the AGN to be $2.6\times10^{11}~L_{\sun}$.\ The
intrinsic AGN luminosity could even be larger if the AGN emission is
not isotropic in the infrared.\ This particular model assumes an
intrinsic AGN luminosity of $3.3\times10^{11}~L_{\sun}$, which was a
free parameter.  So, if this model prediction is correct, $\sim$20\%
of the AGN luminosity is not seen in our line of sight as either
re-emitted or residual AGN energy after absorption.

In this model, the deficiency of the M~82 SED is compensated by the
absorption and re-emission of the AGN light.\ As seen in
Figure~\ref{spmodel2}, the AGN-related emission nicely adds luminosity
to the M~82 SED such that the final SED matches that of the observed
SED.\ The relative contribution to the total observed luminosity is
40\% from the starbursts and 60\% from the AGN at 4.5--120~$\mu$m.  If
we take the intrinsic AGN luminosity ($3.3\times10^{11}~L_{\sun}$)
instead of the observed one (i.e., residual plus re-emitted), the AGN
contribution becomes 65\%.\ Although it is not difficult to reduce the
AGN contribution significantly by adopting a different starburst SED
(e.g., NGC~253), this particular model indicates that it is also
possible to bring up the AGN contribution level to 60-65\% in NGC~6240
despite the absence of any obvious AGN signatures in the mid-infrared
spectrum.\ Given various uncertainties, these numbers are compatible
with the estimates by \citet{Lutz03} based on a number of
AGN-starburst diagnostics including the properties of the mid-infrared
fine structure lines observed by ISOSWS.  The purpose of this
modeling, however, is not to promote the view that the AGN dominates
the luminosity output of NGC~6240 since the model presented here is
just one example in which we pushed the AGN luminosity to its maximum
despite some difficulties.  The intention is rather to illustrate how
subtle the effects can be when the SED of a heavily absorbed AGN (the
residual AGN emission after absorption plus re-emitted components) is
added to that of starbursts.

Perhaps the most serious flaw of this model is the overproduction of
the flux at 13--18~$\mu$m.\ It is difficult to suppress the AGN flux
in this wavelength range because it corresponds to the local minimum
of the extinction curve between the 10 and 18~$\mu$m silicate
absorption.  However, given our limited knowledge of the extinction
curve in this wavelength range, we do not yet consider this problem as
fatal.  Put in another way, if the extinction curve adopted here is
correct, this is the wavelength range in which we may be able to
detect heavily dust-obscured AGNs.

In fact, the choice of an extinction law affects such modeling
substantially.  For the residual AGN spectrum shown in
Figure~\ref{spmodel2}, we used the extinction law by
\citet{Weingartner01} for the Milky Way dust size distribution with
total extinction to reddening $R_{V}=A_{V}/E(B-V)=3.1$ and increased
the amount of extinction between 5 and 9~$\mu$m to fit the extinction
law found for the line of sight to the Galactic center \citep{Lutz96}.
This modification was necessary to suppress the AGN light and produce
a good fit to the observed spectrum.  Without this modification, the
AGN light would have dominated the spectrum at $\lesssim~7~\mu$m with
the assumed level of extinction ($A_{V}=70$~mag).

\subsection{The Limitations of the AGN-Starbursts Power Diagnostics}
When, as seems to be the case with NGC~6240 \citep{Krabbe01}, the AGNs
are highly absorbed in the mid-infrared, the mid-infrared diagnostics
\citep[e.g.,][]{Genzel98,Laurent00,Imanishi00} lose their sensitivity to
AGNs. The most reliable AGN-starbursts power diagnostics are
apparently the hard X-ray--bolometric luminosity correlation for AGNs
and the radio--infrared luminosity correlation for starbursts.
However, neither of these diagnostic relations can be used to estimate
the bolometric luminosity of an AGN or starbursts with an accuracy
better than a factor of a few.  The radio--infrared luminosity
correlation for starbursts is quite tight over a large luminosity
range, but even in the most recent and extensive compilation
\citep{Yun01}, this correlation has an intrinsic 1~$\sigma$ scatter of
almost a factor of two (0.26 in dex).  The scatter is larger with the
correlation between the AGN hard X-ray and bolometric luminosities.
This means that when the starburst and AGN luminosities are comparable
(i.e., within a factor of a few of each other), it is impossible to
determine which dominates and by how much.

The SED variation between the two archetypical starburst galaxies M~82
and NGC~253 (Figure~\ref{spmodelb}) underscores this intrinsic
uncertainty associated with any AGN-starbursts power
diagnostics. Despite the good match of the mid-infrared spectra, their
SEDs differ by a factor of a few in the far-infrared, where most of
the luminosity is coming out.  This would leave room for a substantial
contribution from a heavily dust-obscured AGN that emits mostly in the
far-infrared.

The AGN-starbursts power diagnostics based on mid-infrared data would
likely suffer a larger uncertainty compared with the diagnostics using
hard X-ray and radio data because of the larger effects of extinction
involved, which are difficult to correct accurately.  As seen in
Table~\ref{fratio}, the flux ratio of the two nuclei in the
mid-infrared (7:1) is significantly different from the ratios in the
hard X-ray (3:1) and radio (2:1).  If we take the hard X-ray and radio
ratios as the luminosity ratios of the AGN and starburst components,
respectively, then this means that the the bolometric luminosity ratio
of the two nuclei must be 2--3 regardless of what the relative
starburst/AGN luminosity contributions are. However, for some reason
(most likely due to a larger extinction in the northern nucleus), this
is not the case in the mid-infrared out to 24~$\mu$m.  The
far-infrared power diagnostics (e.g., the \cii\ 158 $\mu$m line
luminosity vs. far-infrared luminosity) are more promising in
principle, and according to these diagnostics, NGC~6240 is a starburst
galaxy \citep{Luhman98,Luhman03}.  However, unless this correlation is
significantly tighter than that of the radio--infrared luminosity
correlation, we will again be left with the factor of a few
uncertainty.

What may be even more problematic is the assumption that the infrared
radiation from a ULIRG is isotropic even when it contains a
dust-obscured AGN.  A natural consequence of a thick dusty torus model
is that most of the AGN radiation is directed perpendicular to the
plane of the torus.  This means that if we are looking at the torus
from an edge-on direction, which may be the case for NGC~6240
considering the large extinction derived from the hard X-ray
observations, most of the AGN luminosity may be invisible to us.  In
other words, even if the observed infrared luminosity is almost all
due to starbursts, this would not necessarily exclude the existence of
an equally (or even more) powerful AGN inferred from the hard X-ray
observations since the true bolometric luminosity of NGC~6240 may be
substantially larger than calculated from our line of sight.
Observationally, the detection of the flat (i.e., unreddened)
reflected hard X-ray continuum \citep{Iwasawa98,Boller03} and a
high-ionization \oiv\ line \citep{Genzel98,Lutz03} suggests that the
AGN is not completely covered up in all directions by Compton-thick
material, which is consistent with a torus-like geometry.

For a toroidal model to work, it also requires the existence of some
material which obscures the AGN in the polar directions of the torus
and softens the AGN radiation. Otherwise, the various emission lines
would show excitation levels higher than observed, which are those of
a LINER in NGC~6240 \citep{Heckman83}.  There are indications in some
Seyfert galaxies that AGN radiation is significantly softened before
reaching the narrow line regions \citep{Alexander99,Alexander00}, but
it remains to be seen if such a model would work in detail to
reproduce the observed properties of NGC~6240.

\section{CONCLUSIONS} 

Using the Keck Telescopes, we have obtained mid-infrared images and
spectra of the luminous infrared galaxy NGC~6240 with sub-arcsecond
spatial resolutions.  The main conclusions are as follows:

\begin{enumerate}

  \item The mid-infrared (and therefore presumably the far-infrared)
  luminosity of NGC~6240 mostly originates from the two nuclei located
  at the positions of the southern and northern radio/hard X-ray
  Sources.

  \item The southern nucleus is especially luminous and compact
  ($<$200 pc in diameter), emitting 80-90 \% of the mid-infrared
  luminosity of the two nuclei.

  \item The southern nucleus seems to be powered by starbursts.  Its
  mid-infrared spectrum is similar to those of the local starburst
  galaxies M~82 and NGC~253, and its mid-/far-infrared SED is broadly
  consistent with those of the starburst galaxies.

  \item At the same time, because of the slight mismatch between the
  NGC~6240 SED and those of the starburst galaxies, a significant AGN
  contribution (up to $\sim$60 \%) to the bolometric luminosity is
  also possible if the AGN is heavily obscured and emits mostly in the
  far-infrared.

  \item Due to the intrinsic variation of properties among a given
  galaxy population, there is a scatter in the relationships inherent
  even with the most reliable and accurate AGN-starbursts power
  diagnostics. When the starburst and AGN luminosities are within a
  factor of a few of each other, this makes it impossible to determine
  in individual galaxies whether an AGN or starbursts dominates the
  luminosity.

\end{enumerate}
 
\acknowledgments

We thank E.\ Sturm, A.\ S.\ Evans, and H.~W.~W.~Spoon for providing us
the ISOSWS spectra, HST/NICMOS images, and the ISOCAM CVF spectrum,
respectively.  We would also like to thank the anonymous referee for
very helpful comments.  The authors wish to recognize and acknowledge
the very significant cultural role and reverence that the summit of
Mauna Kea has always had within the indigenous Hawaiian community.  We
are most fortunate to have the opportunity to conduct observations
from this mountain.  This research has made use of the NASA/IPAC
Extragalactic Database (NED), which is operated by the Jet Propulsion
Laboratory, California Institute of Technology, under contract with
the National Aeronautics and Space Administration.

\clearpage

\input{tab1.tex}

\clearpage

\input{tab2.tex}

\clearpage

\end{document}

%% file: tab1.tex
\begin{deluxetable}{rcrrr}
\tablewidth{0pt}
\tablecaption{N6240 mid-infrared continuum flux densities\label{flux}}
\tablehead{
\colhead{Wavelength} & \colhead{T$_{\rm int}$} & \colhead{Total\tablenotemark{a}} 
& \colhead{South\tablenotemark{b}} & \colhead{North\tablenotemark{b}} \\
\colhead{($\mu$m)}   & \colhead{(s)}  &\colhead{(mJy)} & \colhead{(mJy)} & 
\colhead{(mJy)} 
}
\startdata
 7.9 & 360 &  494$\pm$12  &  275$\pm$3  &  59$\pm$3  \\
 8.8 & 360 &  243$\pm$6   &  142$\pm$1  &  29$\pm$1  \\
10.3 & 315 &  110$\pm$8   &   67$\pm$2  &   7$\pm$2  \\
11.7 & 360 &  316$\pm$4   &  201$\pm$1  &  41$\pm$1  \\
12.5 & 360 &  480$\pm$6   &  375$\pm$2  &  58$\pm$2  \\
17.9 & 480 & 1098$\pm$30  &  919$\pm$9  & 119$\pm$9  \\
24.5 & 432 & 3263$\pm$112 & 2734$\pm$31 & 384$\pm$31 \\
\enddata

\tablenotetext{a}{Measured with a 4\arcsec\ diameter circular beam
 centered between the two nuclei.}
\tablenotetext{b}{Measured with a 1\arcsec\ diameter circular beam
 centered on the nucleus.}
\tablecomments{The uncertainties listed in the table are statistical
  only.  Photometric uncertainties in the MIRLIN data are $\pm$5\% for
  $<20$ $\mu$m and $\pm$10\% for $>20$ $\mu$m.}

\end{deluxetable}

%% file: tab2.tex
\begin{deluxetable}{lcc}
\tablecolumns{3}
\tabletypesize{\footnotesize}
\tablewidth{0pt}
\tablecaption{N6240 continuum flux ratios\label{fratio}}
\tablehead{
\colhead{Band} & \colhead{Flux ratio} & \colhead{Ref}
}
\startdata
~~~~~0.2-10 keV                  & 2.8\tablenotemark{a} & 1 \\
~~~~~1.1~$\mu$m                  & 2.7 & 2 \\
~~~~~1.6~$\mu$m                  & 3.0 & 2 \\
~~~~~2.2~$\mu$m                  & 3.5 & 2 \\
~~~~~7.9~$\mu$m                  & $4.7 \pm 0.2$ & 3 \\
~~~~~8.8~$\mu$m                  & $4.9 \pm 0.2$ & 3 \\
~~~~~10.3~$\mu$m                 & $> 9.6$\tablenotemark{b} & 3 \\
~~~~~11.7~$\mu$m                 & $4.9 \pm 0.1$ & 3 \\
~~~~~12.5~$\mu$m                 & $6.5 \pm 0.2$ & 3 \\
~~~~~17.9~$\mu$m                 & $7.7 \pm 0.6$   & 3 \\
~~~~~24.5~$\mu$m                 & $7.1 \pm 0.6$   & 3 \\
~~~~~1.3 mm                      & $3.2 \pm 0.3$   & 4 \\
~~~~~2 cm                        & $2.5 \pm 0.5$   & 5 \\
~~~~~3.6 cm                      & $2.2 \pm 0.2$ & 6 \\
~~~~~6 cm                        & 2.3 & 7 \\
~~~~~20 cm                       & 2.1 & 7 
\enddata
\tablerefs{(1) \citet{Komossa03}; (2) \citet{Scoville00}; (3) This work;
(4) \citet{Tacconi99}; (5) \citet{Carral90}; (6) \citet{Colbert94}; 
(7) \citet{Beswick01}}
\tablenotetext{a}{Corrected for absorption.}
\tablenotetext{b}{We treat the 3.5 $\sigma$ detection of the northern
  nucleus at 10.3 $\mu$m (Table~\ref{flux}) as an upper limit.}

\end{deluxetable}